**How are cities pledging net zero? A computational approach to analyzing subnational climate strategies**

Authors: Siddharth Sachdeva[1,5], Angel Hsu[1,2*], Ian French[3], and Elwin Lim[4]

[1] Data-Driven EnviroLab, University of North Carolina-Chapel Hill, 131 S. Columbia Street, Chapel Hill, NC 27516
[2] Department University of North Carolina-Chapel Hill, 131 S. Columbia Street, Chapel Hill, NC 27516
[3] Yale-NUS College, 16 College Ave W, Singapore, 138527
[4] Yale School of Environment, 195 Prospect Street, New Haven, CT 06511
[5] University of Chicago, 5801 S Ellis Ave, Chicago, IL 60637
[*] corresponding author

## Abstract

Cities have become primary actors on climate change and are increasingly setting goals aimed at net-zero emissions. The rapid proliferation of subnational governments "racing to zero" emissions and articulating their own climate mitigation plans warrants closer examination to understand how these actors intend to meet these goals. The scattered, incomplete and heterogeneous nature of city climate policy documents, however, has made their systemic analysis challenging. We analyze 318 climate action documents from cities that have pledged net-zero targets or joined a transnational climate initiative with this goal using machine learning-based natural language processing (NLP) techniques. We use these approaches to accomplish two primary goals: 1) determine text patterns that predict "ambitious" net-zero targets, where we define an ambitious target as one that encompasses a subnational government's economy-wide emissions; and 2) perform a sectoral analysis to identify patterns and trade-offs in climate action themes (i.e., land-use, industry, buildings, etc.). We find that cities that have defined ambitious climate actions tend to emphasize quantitative metrics and specific high-emitting sectors in their plans, supported by mentions of governance and citizen participation. Cities predominantly emphasize energy-related actions in their plans, particularly in the buildings, transport and heating sectors, but often at the expense of other sectors, including land-use and climate impacts. The method presented in this paper provides a replicable, scalable approach to analyzing climate action plans and a first step towards facilitating cross-city learning.

## Introduction

The Intergovernmental Panel on Climate Change (IPCC) published the Special Report on Warming of 1.5°C in 2018 declaring an urgent need for global decarbonization to avoid the most dangerous impacts of climate change. Limiting long-term global warming to 1.5°C will require global $CO_2$ emissions to fall by 45 percent from 2010 levels by 2030 and reach net zero around 2050 (IPCC, 2018). Global net-zero emissions is achieved when anthropogenic GHG removals equals or exceeds emissions, customarily measured in units of carbon dioxide equivalents ($CO_2e$). Reaching net zero will require transformative changes throughout global political, social and technological systems, including shifting to renewable forms of energy production; electrifying and decarbonizing energy use in transportation, industry, and buildings; and reducing waste in the food system (IPCC, 2018). Major greenhouse gas-emitting nations, such as the Russian Federation and Brazil, are not on track to meet the goals set forth in their Nationally Determined Contributions (NDCs), meaning that the Paris Agreement is not on track to limit global warming to 1.5°C. Ambitious, voluntary subnational net-zero targets that go beyond national government efforts could help fill the gap in climate ambition. In Australia, for instance, all eight states have pledged mid-century decarbonization targets long before the national government of Australia produced its own commitment (Data-Driven EnviroLab & NewClimate Institute, 2020). Subnational governments' net-zero targets could play a vital role in addressing the climate crisis.

Subnational governments often set net-zero targets through networks that collate individual pledges, such as the UNFCCC Race to Zero campaign (*Race To Zero Campaign | UNFCCC*, n.d.) and the Carbon Neutral Cities Alliance (Carbon Neutral Cities Alliance, 2017). These targets represent a wide range of geographic, socio-political and economic contexts, from megacities with populations larger than 10 million, such as Chengdu and Mumbai, to hundreds of smaller cities and regions on six continents. Public attention on net-zero commitments has increased since 2018, coinciding with key global climate action events, such as the 2018 Global Climate Action Summit, and net-zero announcements by major governments and businesses, such as the UK's net-zero target announcement in July 2019 and Microsoft's carbon negative target announcement in January 2020 (NewClimate Institute & Data-Driven EnviroLab, 2020). Several recent reports have documented the landscape of net-zero target setting. The Data-Driven EnviroLab and NewClimate conducted one analysis that found at least 826 cities and 103 regional governments have set some form of net-zero target, all of which focus on a particular sector or scope of emissions (i.e., direct emissions) (NewClimate Institute & Data-Driven EnviroLab, 2020). In their 2021 analysis, the Oxford Net Zero and the Energy and Climate Intelligence Unit found that 9 percent of all regions in the top twenty-five emitting countries, as well as 13 percent of all major cities with a population over 500,000, have set a net-zero target of some kind (Black et al., 2021). The report notes, however, that a much smaller number of actors have set definitively robust net-zero targets - those aligning with the UNFCCC Race to Zero Campaign's "starting line" criteria of "Pledge, Plan, Proceed, and Publish"(Black et al., 2021). These findings fuel ongoing debates amongst diverse stakeholders concerning the most effective strategies for crafting robust net zero targets.

The number, scale and scope of subnational net-zero pledges has grown precipitously, yet there has been little analysis focused on the strategies and policies local governments are employing to achieve these goals. There are studies analyzing the net-zero targets of specific cities, such as Copenhagen, or specific sectors, such as buildings, yet broader assessments have been limited by data availability and inconsistent documentation (Damsø et al., 2017; Feng et al., 2019). Subnational governments set net-zero targets in an array of various contexts, meaning every target is distinct from the rest, with differences in scope, the use of interim benchmarks, and how pledges are encoded into legislation. A comprehensive, global analysis of net-zero targets would shed light on the strategies subnational governments are employing in their net-zero action plans, the drivers of net-zero targets, and what factors make these policies more impactful.

In this paper, we collected climate policy strategy documents from 318 cities and regional governments (full list in Table S3) that have either declared a net-zero emissions reduction target or have signed on to an initiative that commits them to an equivalent target. Many subnational governments have published planning documents associated with their net-zero targets. These documents typically provide information on the quantifiable emission reduction target (i.e., percentage emissions reduction from a baseline to a target year), specific sectors a city plans to prioritize in their decarbonization effort, policies and strategies to reach their goals, and the timeline of the implementation of policies. We are limited to cities that have published climate action policy and strategy documents, meaning that this sample is not comprehensive, yet it provides insight into how city governments articulate plans to implement their emission reduction targets. We utilize automated natural language processing (NLP) techniques (Gentzkow et al., 2019) to examine patterns and trends in cities' climate strategies and reveal drivers of ambitious climate actions (e.g. economy-wide net-zero targets).

**Complexities in defining net-zero climate pledges**

As more attention is given to the concept of decarbonization (IPCC, 2018), a debate has ensued over what "net-zero emissions" means in practice (NewClimate Institute & Data-Driven EnviroLab, 2020). More than 130 national governments have adopted the goal of "net-zero emissions" (*Net Zero Tracker* , 2021) and a growing number of cities, regions, and corporate actors have made net-zero pledges, yet the concept does not yet have real-world definition (*Race To Zero Campaign | UNFCCC*, n.d.). Article 4 of the Paris Agreement outlines the idea, stating that to limit global warming to 1.5°C, "a balance between anthropogenic emissions by sources and removals by sinks of greenhouse gases in the second half of this century" (Haites et al., n.d.; (UNFCCC, 2015)Haites, Yamin and Höhne, 2013). The IPCC's Special Report on Global Warming of 1.5°C explains that all pathways resulting in less than 1.5°C of warming would require global net-zero carbon dioxide emissions by around 2050, and net-zero greenhouse gas emissions around 2070. It further specifies that even limiting global warming to 2°C - which carries higher risks for crossing irreversible tipping points across the biosphere - means reaching global net-zero carbon dioxide emissions by around 2070.

There is no standard universal definition, however, for what net-zero means when applied to individual actors. A survey of the nomenclature applied to efforts to align targets towards the 1.5°C goal revealed varied terms, including "carbon neutral," "climate neutral," "zero emissions," "climate positive," among others (NewClimate Institute & Data-Driven EnviroLab, 2020).[1] Cities and other subnational governments are left to determine their own definitions of net-zero emissions, and the resulting climate pledges are heterogeneous, often incomparable, and inscrutable to a public audience attempting to make sense of them. Some organizations suggest that actors should focus on reducing their own emissions first and foremost (Bettin et al., 2020)(Carbone 4, 2020), while others allow for the use of offsets, such as renewable energy credits generated outside of an actor's jurisdiction or land-based reforestation credits (C40 Cities for Climate Leadership et al., 2019). The Carbon Neutral Cities Alliance, for example, asks its members to achieve net-zero emissions by reducing total greenhouse gas emissions by at least 80 percent by 2050 (CNCA, 2017). Likewise, some net-zero targets focus exclusively on $CO_2$, while others include other greenhouse gases such as methane or nitrous oxide.

Targets often differ in the breadth of their scopes—some apply to specific sectors, such as buildings, while others apply to all emissions directly produced within a city's jurisdiction. Some cities, such as Seattle, Vancouver, and London, have tried to account for consumption-based emissions—the "indirect" GHG emissions generated from city residents' consumption of goods and services like food, clothing, or electronics—alongside sector-based emissions inventories that track emissions produced within the city from energy use in vehicles, homes, and other buildings, and from waste (C40 Cities for Climate Leadership et al., 2019). The wide variation amongst targets is painted over by the use of "net zero" and other ill-defined blanket terms. It is therefore difficult to determine the ambition of individual actors or to accurately gauge the long-term impacts of their climate policies from net-zero targets alone.

## Methods

Social scientists have, over the past decade, increasingly used Automated Natural Language Processing (NLP) methods that rely on machine learning algorithms to systematically analyze content of text data (Gentzkow et al., 2019; Grimmer & Stewart, 2013; Schwartz & Ungar, 2015). Sourcing quantitative data from text documents, researchers are able to apply automated techniques for both prediction and causal inference. These techniques can be grouped into two categories: supervised approaches, where a researcher specifies a priori a conceptual framework to analyze a text corpus; and unsupervised approaches, where an algorithm identifies patterns and trends in the data (Lucas et al., 2015). Examples of the latter include topic modeling, which has been used to examine linkages between national Paris Agreement pledges and non-state and subnational actors (Hsu et al., 2019) as well as similarities among climate actions of cities, regional governments, companies and national actors (Hsu & Rauber, 2021).

For this study, we apply NLP approaches to accomplish two primary goals: 1) determine text patterns that predict "ambitious" net-zero targets, where we define an ambitious target as one that encompasses a subnational government's economy-wide emissions (i.e., do certain language or textual patterns distinguish between cities that set economy-wide targets and those that do not); and 2) perform a sectoral analysis to identify patterns and trade-offs in climate action themes (i.e., land-use, industry, buildings, etc.). This analysis will help us compare and contrast

[1] Refer to Table 2 in NewClimate Institute & Data-Driven EnviroLab (2020) for a breakdown of a full list of these terms, suggested definitions and how they are used in practice

the ways cities address different emissions sources and examine patterns among subnational governments' climate strategies.

## Preprocessing

In total, climate policy and planning documents from 318 cities were collected between May and July 2020 by conducting Internet searches of each city's name plus keywords including, "climate action plan," "net-zero plan," and "carbon neutral plan." Some subnational actors have produced multiple reports or plans detailing their proposed climate actions;  the collected documents represent 318 unique cities, which are predominantly located in North America (*n*=78) and Europe (*n*=202) (see Figure S1).

The format of these documents varies substantially, with lengths ranging from a few pages, such as the city of Fremont's Resolution No. 2019-03, to Barcelona's 164-page climate action plan that lays out a roadmap for climate action to 2030 (Barcelona City Council Area of Urban Ecology, 2018). The content of these documents also varies substantially, exhibiting the aforementioned diversity generally observed among net zero targets. The plans use a wide variety of terms to describe their targets including, for instance, Cape Town's commitment to "carbon neutrality" by 2050 and Tokyo's promise of a "zero emission" city (Tokyo Metropolitan Government Bureau of the Environment, 2021). The timelines laid out in these documents also vary from plan to plan; Copenhagen, for example, pledges to reach net-zero emissions by as early as 2025, although the majority of actors aim to reach net zero by 2050  (Copenhagen Technical and Environmental Administration, 2016). Some plans focus on a subset of years within larger pathways, such as Portland's Climate Action Plan Progress Report, which charts the city's proposed actions through 2030 along its larger journey to net zero emissions by 2050 (Portland Bureau of Planning and Sustainability, 2017). Some cities had more than one climate policy or strategy document, and so we concatenated these documents together so that every city had one document corpus for the analysis.

Prior to analysis, the raw documents were first converted into a text file through optical character recognition (OCR) using the pytesseract (version 0.3.8) implemented in python (Indravadanbhai Patel et al., 2012), and non-English documents were translated to English using the Google Translate API. Any document that was less than 50 characters in length was excluded from analysis, leaving a total of 318 cities. To pre-process the text data corpus, the text was tokenized (split into words) and lemmatized (converted to root form), and stopwords were removed, including punctuation and numbers, proper nouns, pronouns. Next, the text was converted into a numerical representation suitable for analysis, referred to in NLP methods as text featurization or tf-idf (term-frequency-inverse document frequency) featurization. Tf-idf is a commonly-used NLP method to convert text to a numeric representation that "assumes the importance of a term relative to a document is inversely proportional to the frequency of occurrence of this term in all the documents" (Tan, 1999; Zhang et al., 2011). In other words, commonly appearing words in one text would have a low tf-idf score if it appears frequently in other documents and conversely, a word that appears frequently in one document that does not tend to be common in all documents would have a tf-idf score closer to 1. To calculate tf-idf scores for the text featurization process, we computed word-count vectors of 1-grams (word) and 2-grams (word pairs), filtering terms that are in less than 10 percent of documents, and then normalized the counts by dividing by each term's frequency in the whole corpus. Using this method, we produced a numerical representation that allows us to understand how important a term is to a document corpus.

**Predicting 'ambitious' economy-wide net-zero targets**

In addition to collecting climate policy and strategy documents, we collected publicly-available information detailing cities' quantitative emission reduction targets (see Supplementary Table S1) to identify whether a city has made a pledge to decarbonize or reach net-zero emissions. We coded each city in our database as pledging a net-zero target according to whether the city had signed on to the UNFCCC's Race to Zero (*Race To Zero Campaign | UNFCCC*, n.d.) campaign, which is a coalition of initiatives that commit members to net-zero targets, or if they have specifically mentioned "carbon neutral" or "net zero" as part of their commitment. We also classified subnational actors that commit to at least an 80 percent emissions reduction target as those with an economy-wide net-zero target. In total, 242 out of 318 cities in our database were identified as pledging an economy-wide net-zero target.

We used this binary identifier (i.e., whether a city has pledged an economy-wide net-zero target) to identify patterns by training a logistic regression-based model applied to the tf-idf text features of the climate action plans using the scikit-learn package (Pedregosa et al., 2011). Logistic regression is a common statistical model used in NLP studies to predict the probability of a document that belongs to one of two binary classes (Jurafsky & Martin, 2021). By evaluating the text the model uses to make these predictions, we can uncover statistically significant patterns associated with subnational governments that are ambitious (i.e., have pledged economy-wide net-zero targets) in their climate actions.

Our model specification is in Equation 1 and 2, and our objective function used to fit parameters to the model is represented in Equation 3 and 4.

$$\log\left(\frac{p_i}{1-p_i}\right) = \beta_0 + \sum_j \beta_j X_{ij}$$

$$p_i = P(Y_i = 1)$$

$$\min_{\beta_0,\beta} \lambda\|\beta\|_1 + \sum_{i=1}^{n} c_i \log(e^{(-Y_i(X_i^T\beta+\beta_0))} + 1).$$

$$c_i = \sum_{j=1}^{n} \frac{\mathbb{1}_j}{n}, \mathbb{1}_j = \begin{cases} 1 & Y_j = Y_i \\ 0 & Y_j \neq Y_i \end{cases}$$

- $X_{ij}$ is a matrix of the tf-idf bi-gram term frequency features $j$ for each city $i$;
- $p_i$ is the probability that a given city has an economy wide net zero target.
- $\beta_0$is the intercept, representing the log of the odds for the average city;
- $\beta_j$are coefficients assigned to each term representing the contribution of that term to the prediction of whether a city has an ambitious climate plan or not.
- $Y_i$ for each city $i$ is a binary outcome variable whether a city pledges an economy-wide, net-zero target;
- $\lambda$is the L1 regularization parameter (explained in more detail below);
- $n$ is the total number of cities
- $c_i$ is a reweighting factor we apply to correct for an imbalance in the distribution in the dataset (i.e., 76 percent of the cities in the dataset have an economy-wide net-zero target).

*L1 Regularization*

Since our document corpus has significantly more features (i.e., terms in our vocabulary) than datapoints (i.e., cities), there is risk the logistic regression model overfits the data. To address potential overfitting due to the model complexity, we apply L1 regularization (Tibshirani, 1996), a technique that is frequently applied in machine learning and neural networks that reduces a model's complexity to avoid overfitting (Yuan et al., 2012). L1 regularization forces the model to use a small set of terms to predict net zero targets. It achieves this by applying a penalty to the parameters in this regression to force the model to use only the most predictive terms and ignore the vast majority of the 8,000-word vocabulary features. This step makes the model interpretable, allowing us to identify a small set of terms that best captures what predicts certain plans to be more ambitious than others.

*Model validation*
To test the accuracy of the model, we refit the model on 50 different train-test splits (i.e. randomly sampled subsets of the data), calculated out of sample accuracy using leave-one-out cross-validation, and averaged the out-of-sample performance metrics from each of these fits. We then average the coefficients from these different fits to ensure that the resulting predictors are robust to sampling error. Finally, we generated p-values for each of the terms used by our model using a chi-square test of independence. The results of the model performance evaluation are detailed in Supplementary Table S2. We report accuracy metrics ranging from 0 to 1, including binary f1-score, precision, and recall averaged over 50 out-of-sample test sets. Because out of sample binary classification accuracy is the primary metric by which we evaluate logistic regression, these results indicate that the model we learn is robust to unseen data and therefore useful for interpretive analyses of the coefficients.

**Key term-based topic analysis**
To examine common themes among these cities' climate strategies, we developed keyword lexicons) (Supplementary Table S3) associated with primary greenhouse gas-emitting sectors where cities are likely to be pursuing emissions reductions. We started by defining this set of nine sectors based on the most common emission sources for cities (Intergovernmental Panel on Climate Change & Intergovernmental Panel on Climate Change, 2015), as well as the most probable topics and sectors for cities identified in Hsu and Rauber's (2021) text analysis of city, region, company, and country climate actions. These topics include:

1. Land use
2. Industry
3. Buildings
4. Transportation
5. energy/electricity
6. Heating
7. Waste/pollution
8. Climate Impacts
9. Offsets

We then gave each of these nine categories seed terms, which we used to find other similar terms for each category. We accomplished this using using word2vec similarity (Honnibal & Montani, 2017) implemented using the *spacy* package in python, which is a computational measure of similarity between words based on how they're used in a large corpus (Mikolov et al., 2013). Combining a high quality set of seed terms inspired by the prior work on cities' biggest emissions sources with this word2vec similarity search allows us to manually curate a large set of representative keyterms for each topic. This approach was inspired by dictionary-based methods (as described in Gentzkow et al., 2019), where a researcher manually curates terms based on domain knowledge and uses those terms to form numerical representations of text that highlight features the social scientist cares about. In our case this process produces manually curated lexicons aided by word vector models that can detect discussion of different emissions sources, allowing us to determine the topics of interest. We then construct a topic vector for every climate plan with nine topics each, where every element of the vector is the number of terms in a plan associated with a certain topic, thereby producing a topic-based representation of the climate plans.

We used these vectors to delineate patterns associated with how these topics are discussed in the net-zero plans. We explore these associations in three ways. To first understand how much the average city is talking about each topic, we measure the median topic count (i.e., how many words belonging to each topic is mentioned) over all cities. Second, we use factor analysis (Costello et al., 2005), a statistical dimension reduction technique used to compress variability among observations with many variables into a smaller number of unobserved variables called latent factors. These latent factors are based on groupings of words that can reveal themes explaining what is driving a city's focus on certain scopes of climate issues. Finally, we plot the distribution of cities along these latent factors to understand the level of variability in the strategies cities are taking.

**Results**

**Summary Statistics**

| Statistic | N | Mean | St. Dev. | Min | Pctl(25) | Pctl(75) | Max |
|---|---|---|---|---|---|---|---|
| population | 315 | 429,236.000 | 1,302,188.000 | 88.000 | 5,620.500 | 252,001.000 | 13,929,286.000 |
| baseline_year | 273 | 2,005.538 | 7.220 | 1,990.000 | 2,005.000 | 2,010.000 | 2,017.000 |
| percent_reduction | 298 | 70.382 | 40.209 | 0.000 | 35.500 | 100.000 | 504.000 |
| emis_per_capita | 255 | 6.372 | 5.058 | 0.002 | 3.471 | 7.246 | 35.140 |

Table 1. Summary statistics of cities included in the analysis.

***Predicting economy-wide net-zero targets***

Our logistic regression yielded predictors that help us identify traits common among ambitious net-zero plans. We find that cities that have set economy-wide net-zero targets have significantly different ($p < 0.01$) language use patterns than cities that do not. The top predictive terms of whether a city pledges a net-zero target are identified in Figure 1, which shows the log(coefficients) for terms that are most highly associated with cities that pledge economy-wide net-zero targets. We group these terms into four descriptive categories or themes to facilitate interpretation: 1) specific, quantitative metrics; 2) identifying emission reduction sources; 3) governance, and 4) human-centric approaches. Mining cities' climate policy documents for specific mentions of the predictive terms allowed us to better understand the context surrounding each term and common patterns in their usage across cities. We discuss these predictive terms for cities pledging economy-wide net-zero targets (i.e., referred to in this section as "ambitious cities") and their context within our climate action plan corpus.

Theme 1. Specific, quantitative metrics

- Opportunity reduce
- Percent
- Year use
- Square
- Ghg reduction

- emissions generate
- Life cycle

Ambitious cities' climate plans tend to include specific quantitative metrics in their strategies. Mentions of targets specifying the **percent** emission reduction or increases in budgets, space-based measures like **square** feet in the context of building footprint or greenspace added, or mentions of a reference **year used** to understand changes in numbers over time. These cities also identify specific **ghg reduction** goals, discussing the **lifecycle** costs and emissions of specific projects, identification of **opportunities to reduce** emissions, and defining the scope of e**missions generated** within city limits.

Theme 2. Emissions sources

- Wood
- Chp (combined heat and power)
- Passive
- Boiler

We find mentions of actions specifically focused on emission sources tend to be predictive of cities setting economy-wide net-zero targets. Building-related emissions frequently appear as a primary focus sector. In heating, as cities focus on transitioning away from gas **boilers**, increasing the use of **combined heat and power** systems is frequently mentioned, as is encouraging **passive** design of buildings to use solar heating. **Wood** is mentioned in the context of construction material, a carbon sink in forests, and also as a heat source.

Theme 3. Governance

- Mayor
- **Fees**
- **energy plan**

Ambitious cities' reduction targets tend to include or suggest specific governance mechanisms to implement their climate actions. They often include a message from the **mayor** indicating direct buy-in from the leadership of the city. Many mention either reducing **fees** for renewable energy projects or introducing fees for fossil fuels, and they contextualize how the city's energy use fits into a regional or national energy plan.

Theme 4. Human-centered:

- Inclusive
- Advocate

Ambitious cities tend to discuss the interaction of their climate plan with their communities. Their plans also discuss quality of life improvements for residents through **inclusive** public spaces and **inclusive** economic opportunities. These more ambitious plans often set out a plan for engaging directly with their communities, **advocating** for needed changes rather than making mandates. These results indicate that community consideration and engagement might be an important aspect of climate action plans.

***Identifying Key themes in Cities' Climate Action Plans***

The key term-based topic analysis yielded analysis of themes in subnational climate action plans. Overall, we find that cities most frequently emphasize climate mitigation measures focused on energy-related actions (Figure 2) compared to other topics we analyzed. The frequency of common themes parallels the largest climate emission sources from cities - energy consumption from buildings and the transportation sectors. Nearly all cities tend to discuss strategies aimed at specific energy sources or renewable energy, as well as actions for the electricity sector. The 'buildings' topic was also identified commonly in cities' plans, including residential building actions and retrofits. Transportation actions focus on increasing pedestrian walkability, as well as mixed modal options, from public transit to cycling. Less commonly mentioned in cities' action plans were the topics representing land-use, heating, pollution/waste, industry, impacts and offsets.

We also observe trade-offs in these topics (Figure 3). Cities that tend to emphasize energy do at the expense of discussing land-use and climate impacts, as the two are negatively-correlated (r=-0.44 and r=0.4, respectively). Similarly, cities that highlight climate impacts are less inclined to focus on buildings (r=-0.36), although the negative association is rather weak. Those that place emphasis on pollution/waste also mention land-use related efforts (r=0.54), and actions on buildings and energy (r=0.38) also go hand in hand.

As expected, these emphasis on certain topics over others varies by city. Individual city wordclouds are provided in Supplementary Table S4. In Figure 3, we show representative wordclouds for Munster and New Bedford, with the sizes of the words corresponding to their tf-idf score, and the color of the word corresponding to their topic membership. From the word clouds we observe Munster's focus on urban infrastructure issues such as parking and building standards, while New Bedford is much more focused on ecological factor issues like ocean stewardship and watersheds. Munster is highly focused on transportation-related emissions while New Bedford dedicates a significant portion of its plan to climate resilience, such as addressing climate impacts through habitat restoration and green infrastructure.

*Factor analysis of prominent city climate action themes*

From these nine commonly-emphasized topics in cities' climate action plans, we further reduced the complexity and heterogeneity of cities' strategies into two dominant themes, which we labeled "Ecology" and "Infrastructure" after inspecting which of the nine topics were most strongly associated with each factor. As Figure 5 illustrates, the Ecology factor is associated with topics focused on pollution/waste, land-use, and impacts, while the Infrastructure factor is associated with the heating, building, energy and transportation topics. Between these two factors, we observe trade-offs; topics that correspond to high values in one factor tend to correspond to low values in the other factor, and vice versa. Overall, we find that the median city is more focused on urban infrastructure than ecological factors. We also found outliers for both factors and two kinds of outliers: those cities that emphasize ecological factors more than most cities, yet tend not to describe urban infrastructure-related efforts as much as most cities (Supplementary Figure 2).

By plotting the factor scores of these cities across the “Infrastructure” and “Ecology” factors in four quadrants (Figure 6), we can identify common themes in how certain types of cities plan to tackle climate action. For example, we see that cities in Quadrant II (the top right quadrant) tend to focus somewhat equally on ecological considerations and urban infrastructure approaches to emissions reductions. Cities in this quadrant include a large number of North American cities, such as San Francisco, Austin, Los Angeles, and Chicago, as well as others, such as Singapore and Tokyo. Examining the plans of these cities can help us better understand why they fit into this quadrant. San Francisco, for example, has high values on both the Ecology and Infrastructure factors. Its plan provides detailed discussion of waste and pollution issues, specifically citing commitments to reduce waste by 15% and disposal through landfill by 50% by 2030 as part of its larger goal of achieving net zero emissions by 2050. At the same time, it also mentions consideration for transportation and industry issues, including its goal of achieving complete electrification of all vehicles by 2050 (San Francisco Department of the Environment, 2019).

Cities in the top left quadrant (Quadrant I) tend to have a strong focus on Urban Infrastructure issues, but less consideration for Ecological Considerations. London, for example, falls neatly into this quadrant because its plan extensively focuses on building and energy strategies, but has a low score for considering climate impacts or land use. Indeed, the city’s plan highlights that it aims to achieve netzero emissions mainly through improved efficiency in its transport, building, and energy networks. These topics are also important in determining Philadelphia’s positioning - this city’s plan describes energy and building strategies, but performs relatively poorly in considering land use issues or climate impacts. We see the city’s industrial focus even in the title of its plan: “Powering Our Future: A Clean Energy Vision for Philadelphia.” This document by no means ignores Ecological Considerations and reports that achieving 100% clean energy is critical for meeting the city’s goal of 80% reduction in greenhouse gasses by 2050 - yet it focuses heavily on showing how clean energy will benefit the economy by “creating local jobs, lowering utility bills, and improving air quality for Philadelphians” (Philadelphia Office of Sustainability, 2018). Other cities in Quadrant I with similar economic emphasis in their plans include Munster and Fontanafredda.

Cities in the bottom left quadrant (Quadrant III) tend to have little consideration for either Urban Infrastructure issues or Ecological Considerations as we have classified them. What must be immediately considered is that cities in this quadrant tend to share less detail overall in their plans, typically due to the length or format of their reporting documents. The Belgian municipality of Izegem, for example, delivers merely a one page summary report, while the Chilean capital of Santiago and the Portuguese capital of Lisbon share a eight page report and a twenty-eight slide summary presentation, respectively. It is somewhat surprising to see capital cities among those without detailed plans for achieving net-zero emissions, particularly in 2020 - the year that countries are supposed to ratchet up their commitments.

Cities in the bottom right quadrant (Quadrant IV) tend to have a strong focus on Ecological Considerations and a less prominent focus on Urban Infrastructure. This quadrant includes cities such as Tainan, Vancouver, Durban, Buenos Aires, and Sydney. Looking at Tainan and Vancouver, we can find that cities in this quadrant tend to have plans that have extensive discussion of land use and pollution or waste issues, but neglect the topics of heating and energy. For example, Vancouver's plan is a comprehensive ninety-three page document with ten chapters corresponding to ten environmental goals, but only six pages involve discussion about developing a new "green economy" within the city - most of the other chapters focus on topics such as zero waste or improving access to nature. Not all of these cities have high values for their plans' representation of climate issues specifically - in many cases, a strong focus on land use is what explains their position in this quadrant.

## Discussion

More than 11,000 subnational governments (United Nations Framework Convention on Climate Change (UNFCCC), 2021) are pledging voluntary climate action, and more cities are expected to follow. Scientific and political calls for global decarbonization and individual actors to "race" (to establish their own plans and strategies to align climate efforts to meet these goals are further driving more participation. Alongside these efforts are increasing demands for greater accountability of non-state actors, including subnational governments, to demonstrate credible actions and progress to meet their climate goals (UN News, 2021). A critical component to substantiating these pledges will be more scrutiny of cities' climate policies and strategies, which is a challenging task given the heterogeneity of their reported formats, languages, and availability (Hsu & Rauber, 2021).

These climate action plans require integrated approaches to climate change mitigation, although urban climate researchers have identified challenges in doing so due to a range of knowledge gaps. For instance, Mi et al. (2019) point to a lack of understanding the roles of urban sectors in climate change mitigation as well as a gap in knowing how cities choose climate mitigation strategies and local actions prevent such planning (Mi et al., 2019). Although our study does not delve into the motivations and decisionmaking behind cities' climate action planning, it does provide a landscape analysis of common themes and patterns in existing climate action strategies for cities to begin to identify potential opportunities for learning (see Neij & Heiskanen, 2021). A primary advantage of the machine-learning based natural language processing (NL) approach is achieving a replicable, scalable method for broad pattern and theme analysis of heterogeneous, diverse climate action plans as a starting point for further, detailed examination. We discuss some of the key findings and their implications for the broader study of subnational climate actions below.

### *Common characteristics in 'ambitious' climate action plans point to specificity and governance*

"Ambitious" climate action plans - quantified, economy-wide net-zero emissions target of 80 percent or higher - tend to emphasize specific metrics and targeted sectoral actions. While this conclusion may seem obvious, historically the connection between urban emissions inventories and climate action plans has been less certain. (Boswell et al., 2010) note that only about 80 cities in the United States at the time had developed climate action plans based on greenhouse gas emissions inventories, echoing a gap (Yalçın et al., 2012) noted in their analysis of European municipal or 'territorial' climate action plans. One of the main challenges for cities in developing action plans is determining long-term quantified mitigation targets, which lead municipal authorities to frequently adopt national, regional or international targets for their climate plans (Yalcin and Lefevre, 2012). Two challenges emerge with respect to this phenomena - targets are frequently set without any understanding of whether a city has the capacity to meet them; or second and conversely, cities could actually achieve higher ambition.

Although we found that mentions of specific quantitative indicators, such as specific emission reduction targets, reference years and target years, among other metrics, were predictive of cities setting net-zero or equivalent emission reduction targets, this result does not necessarily imply that the cities will take the necessary implementation measures to achieve these goals. Hsu et al., (2020) found in an evaluation of just over 1,000 cities participating in the European Covenant of Mayors for Climate and Energy that setting more ambitious, longer-term emission reduction targets were weakly negatively or barely associated with performance. Our finding, therefore, that governance-related language, such as the involvement of citizens in addition to top-level leadership from mayors, were also predictive of ambitious, net-zero equivalent targets. In fact, we found language and terminology related to the word 'mayor' to be the strongest predictor of whether a city has pledged a net-zero or equivalent emission reduction goal. Through manual inspection of the individual strategies themselves, we find that some of the ambitious cities' plans feature forwards or introductory remarks from the city mayors themselves or specific mention of mayoral support and buy-in. This trend is consistent with previous studies that have found that "active local politicians" and "a supportive local community" are key enabling factors of ambitious climate policies. It also echoes other research that find municipal climate action plans tend to emphasize self-governance as a means of enacting climate policy (Bulkeley & Kern, 2006; Croci et al., 2017), although there are questions in terms of its effectiveness (Hoppe et al., 2016).

*Dominant emphasis on energy-related emission sources and actions could overlook key mitigation opportunities*

Our results found that cities are predominantly focused on the energy sector in their climate action plans. Since energy-related emissions from stationary energy sources (e.g., fuel combustion and electricity use in residential and municipal or commercial buildings) comprise the majority or more of cities' greenhouse gas emissions (Wei et al., 2021), it is unsurprising we found energy-related climate actions to be the most commonly discussed sectors across our sample. The emphasis on these energy-related emissions, however, could overlook some important opportunities for emissions reductions. As (Yalçın et al., 2012) note, "local climate mitigation initiatives often focus on win-win measures in the field of energy saving," and may be missing some key sectors responsible for local emissions, such as consumption-based or Scope 3 emissions. (Harris et al., 2020) found that when accounting for consumption-based emissions in 10 European cities, these emission sources were twice as large as production-based emissions that cities typically report and base their climate action plans on. Fewer than 10 percent of the cities we analyzed specifically mentioned 'Scope 3' emissions in their climate action plans. Cities falling in Quadrant III of our factor analysis, articulating climate strategies that do not heavily emphasize language in the Ecology or Infrastructure themes may warrant further investigation to understand whether they may be focusing on some of these critical gaps that other cities are not.

*Trade-offs in topical foci limit opportunities for cross-sectoral synergies*

Our observation of trade-offs in certain topics in cities' climate actions further reinforces cities' tendency to silo climate actions rather than opt for integrated approaches. Cities climate action plans predominantly emphasize language and strategies related to infrastructure often at the expense of 'ecology'-related strategies. Our finding that energy and building-related activities are discussed at a cost of climate impacts suggests cities are missing key opportunities to articulate mitigation actions in tandem with adaptation. (Lee et al., 2020) suggest that cities addressing both mitigation and adaptation together can capitalize on shared capacity building, more efficiently utilizing limited resources and taking advantage of synergies between the two, arguably equally important, climate concerns for cities. As (Otto et al., 2021) note, city officials often treat mitigation and adaptation separately, developing distinct strategies and seldomly considering their integration.

The trade-off observed between the land-use and energy and transport topics suggests cities are also falling short of integrating spatial planning as a critical tool for climate mitigation (Zanon & Verones, 2013). For example, infusing climate planning with spatial considerations can result in deeper greenhouse gas emissions reductions since it can affect transportation demand and overall energy demand (Neij & Heiskanen, 2021). The opportunity for cities to bridge spatial planning in their climate action plans is an opportunity urban climate scholars point to as a key mode for accelerating innovation (Broto, 2017).

*Limitations*

The main limitation of our study is in terms of regional diversity - most of the cities that have published climate action plans are located predominantly in the Global North. This gap is one that has been echoed in previous studies of subnational climate action (Angel Hsu et al., 2020; Kuramochi, 2020), given the longer experience of cities in Europe and North America in participating in transnational climate initiatives and developing climate action plans (Boswell et al., 2010; Yalçın et al., 2012). We are also limited to analyzing cities that have published a climate action strategy or net-zero climate action plan. Since this is a rapidly developing landscape, with as of November 2021 more than 733 cities signed on to the UNFCCC's Race to Zero campaign, we would expect more cities to develop climate action plans, with specific net-zero strategies, in the near future. Another limitation is lacking information regarding the outcomes or implementation of these strategies - simply adopting a emission reduction target or articulating a climate strategy does not necessarily imply a city has taken necessary action to implement them. Our study, therefore, represents an analysis of content rather than outcomes and does not predict whether ambitiously-articulated climate plans result in achievement of results.

Despite these limitations, our study demonstrates the utility of machine-learning aided text analysis techniques in systematically analyzing climate action plans and provides a replicable and scalable approach as more cities join global efforts to set their own net-zero and decarbonization goals.

**Software**

Statistical analyses were conducted using the R statistical programming environment (Version 3.6.2) and in python using the libraries identified in the Methods section.

**Data Availability**
Data compiled for the study is available at UNC dataverse (unc.dataverse.com). Code for analysis is available on GitHub (www.github.com/datadrivenenvirolab).

**Acknowledgements**
This research was supported by a 2018 National University of Singapore Early Career Award (Grant Number NUS_ECRA_FY18_P15) to A. Hsu and IKEA Foundation Award (G-2001-01507). We would like to thank Zhi Yi Yeo, Vasu Namdeo, Yinxi Tan, and Adele Tan of Yale-NUS College and Amy Weinfurter of the WaterNow Alliance for their assistance in data collection.

**Author Contributions**
AH and SS conceived of and designed study. AH conducted data cleaning and statistical analysis. SS conducted the statistical programming and analysis. IF, AH, and SS developed visualizations. All authors contributed to research and writing.

**Competing Interests**
The authors declare no competing interests.

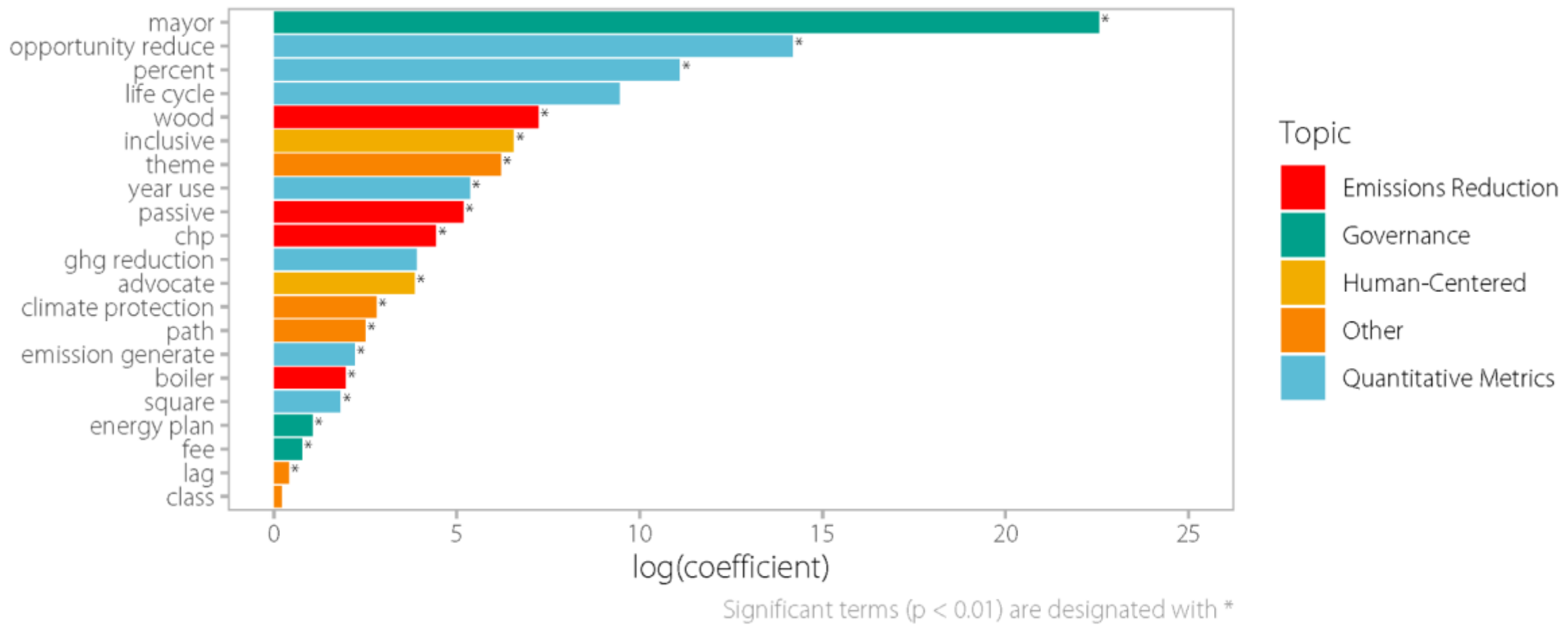

**Figure 1.** Terms predictive of economy-wide net-zero targets. P-values less than 0.01 indicate significant terms.

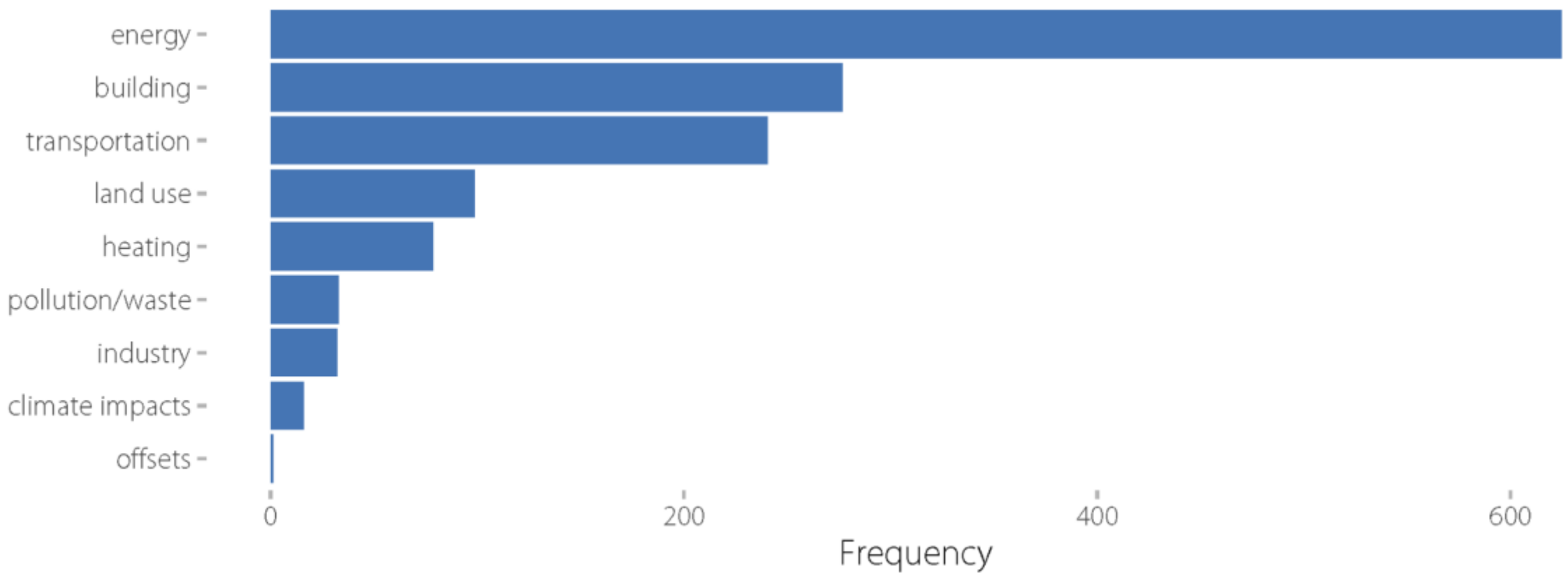

**Figure 2.** Median topic count for cities climate action strategies in sample (n=318).

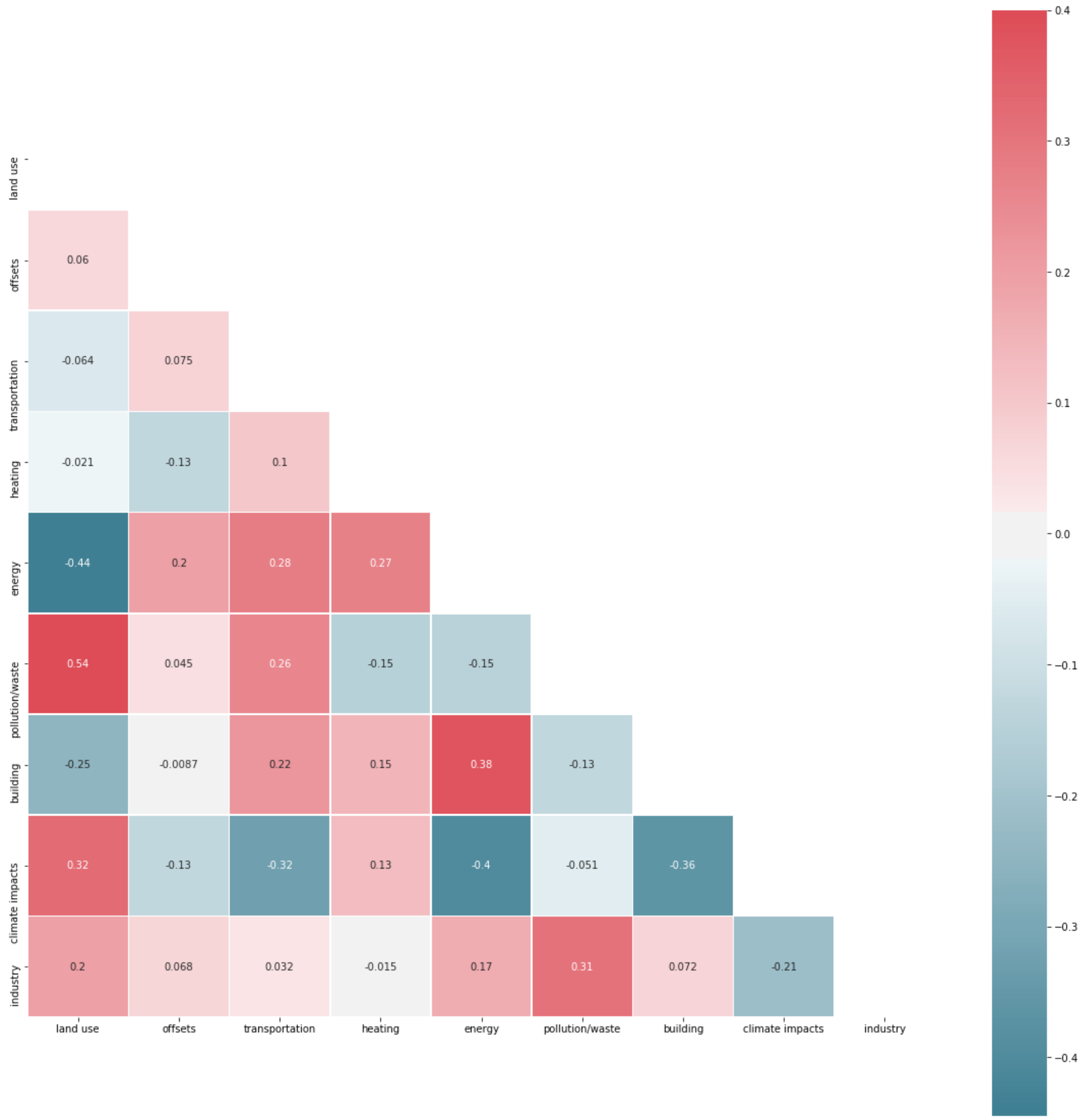

**Figure 3.** Correlation plot for key-term based topics in our corpus.

a) Munster

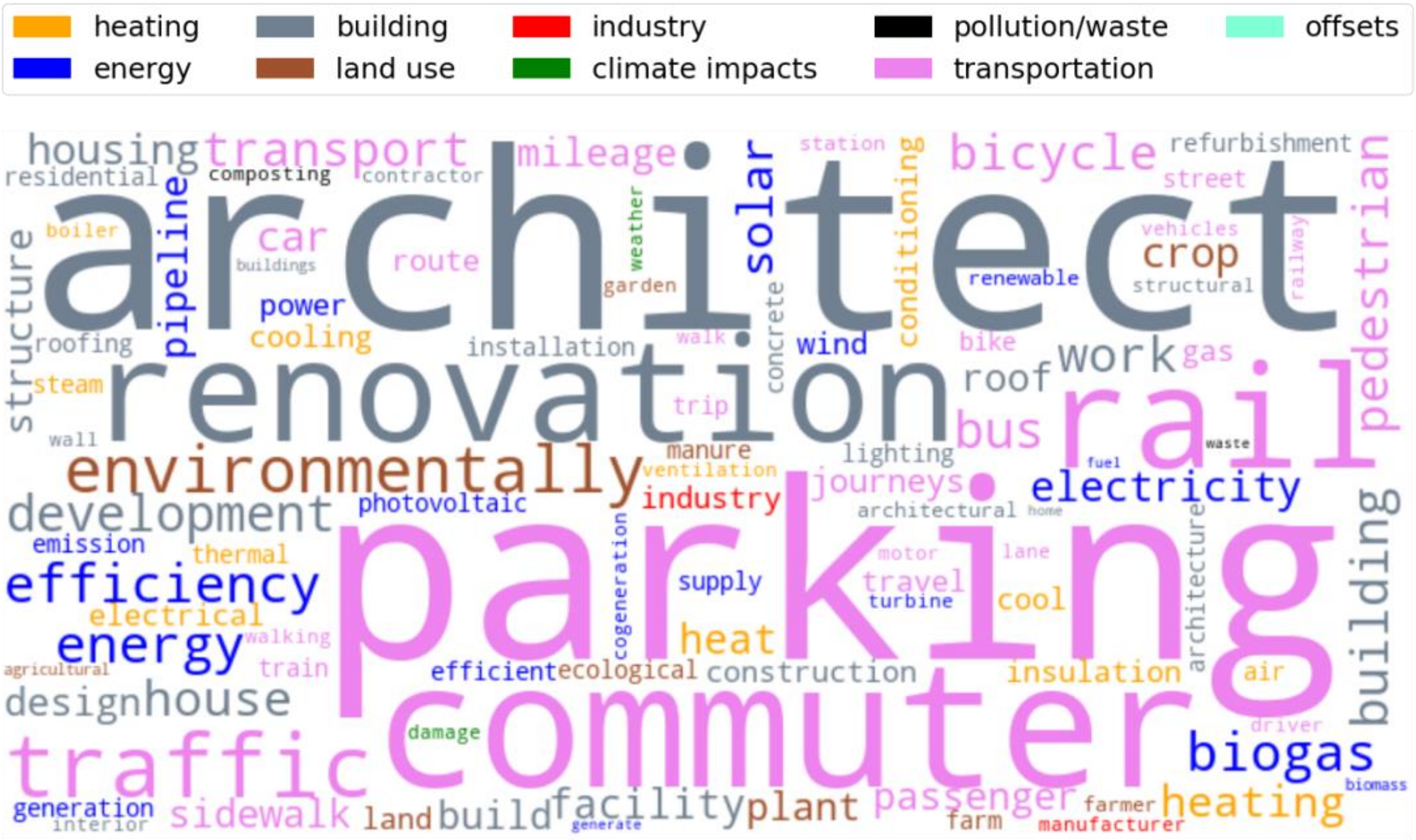

b) New Bedford

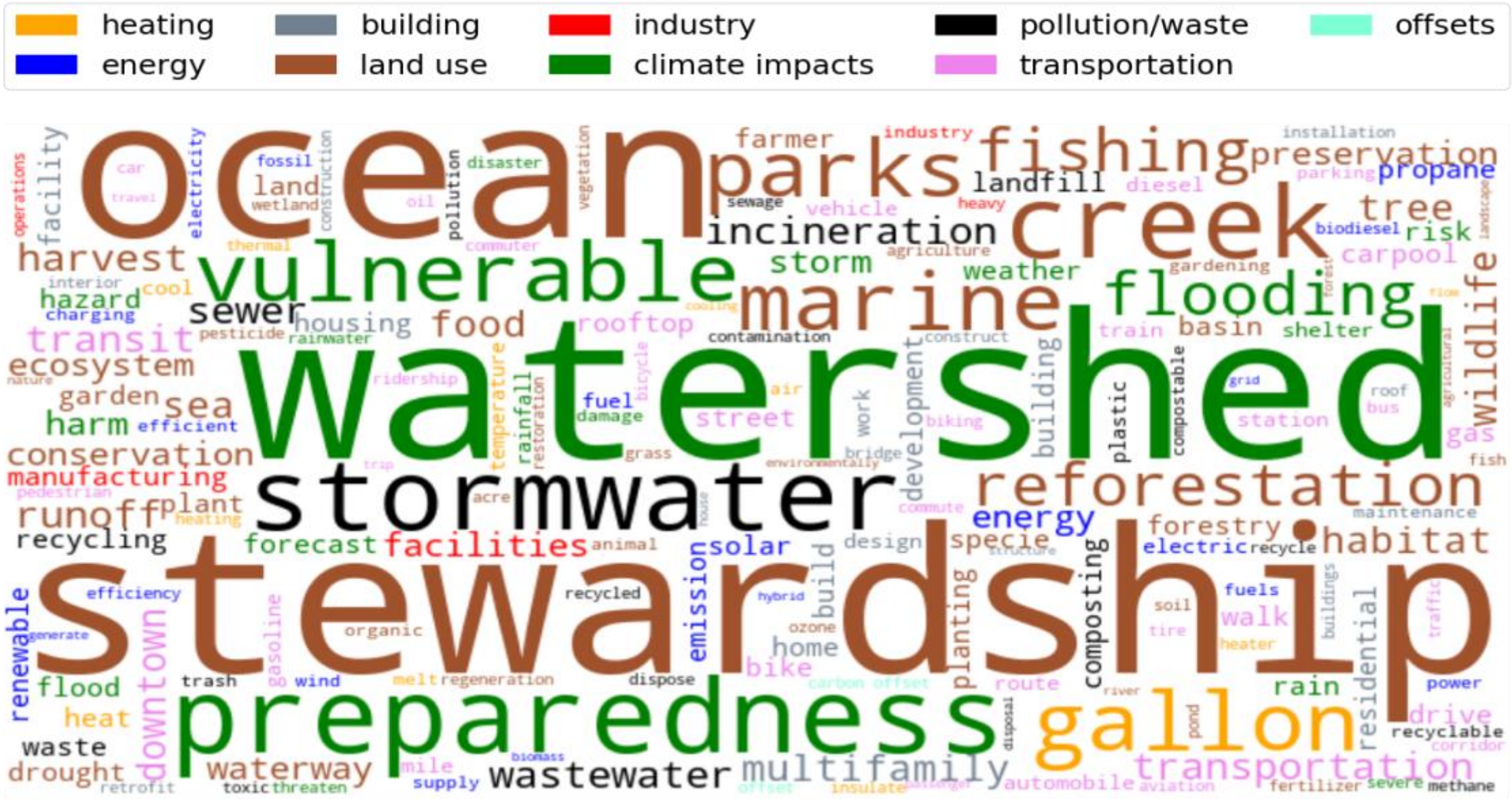

**Figure 4.** Topic word clouds for the cities of a) Munster and b) New Bedford. The size of each word corresponds to its tf-idf score or relative frequency of that term in that cities' climate plan. The color corresponds to the topic associated with each term.

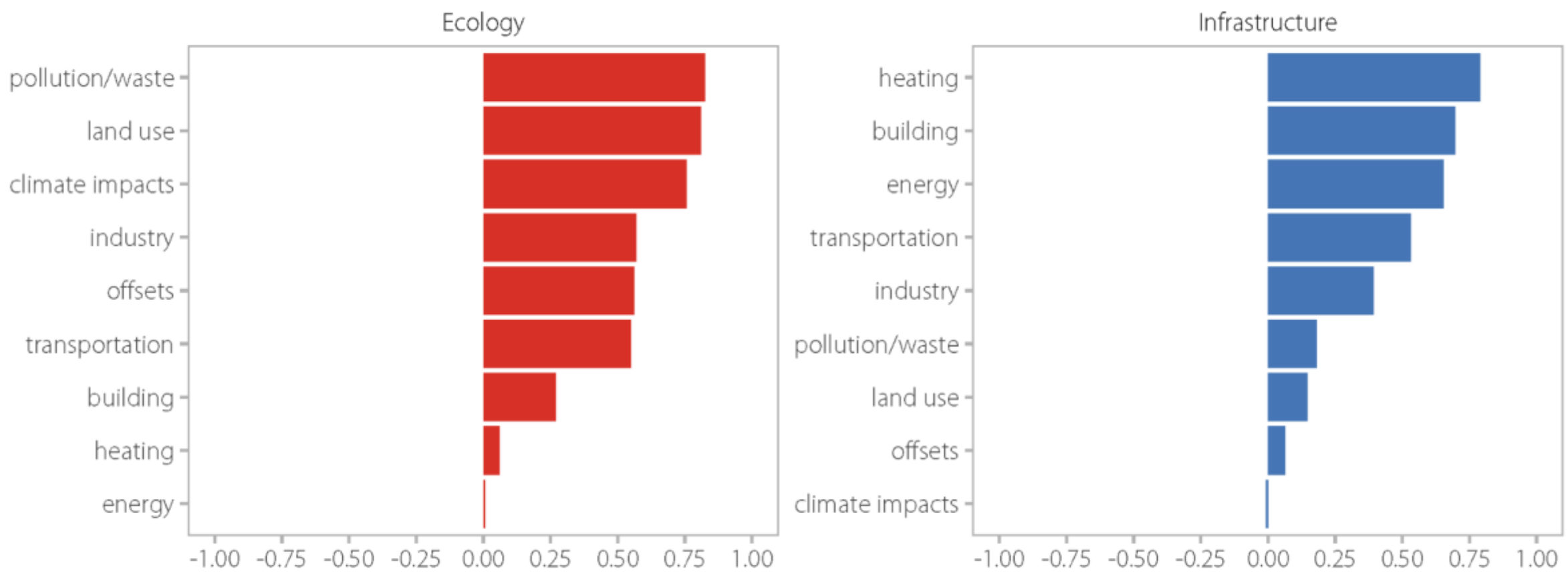

**Figure 5.** Two factors corresponding to Ecology and Infrastructure themes were determined from cities' climate action plans.

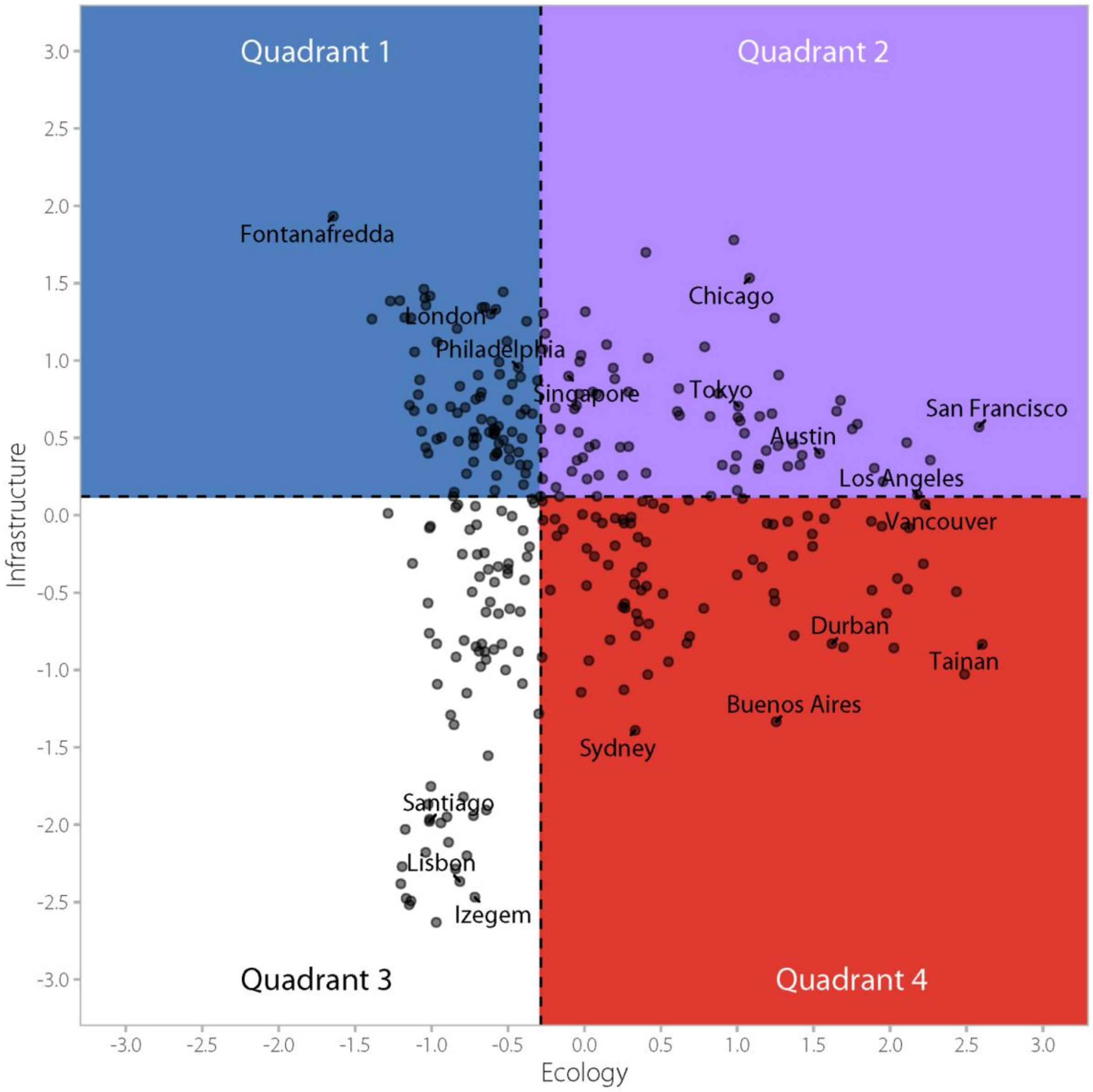

**Figure 6**. A plot of cities' factor scores according to two underlying latent factors: infrastructure and ecology.

# How are cities pledging net zero?A computational approach to analyzing subnational climate strategies

Supplementary Materials

**Supplementary Table 1.** Data sources for city climate action commitments.

| **Climate Action Platform** | **Data source** |
|---|---|
| **C40 Cities for Climate Leadership Group** | C40 Cities for Climate Leadership. Accessed February 2021 from: https://www.c40.org/cities. |
| **ICLEI Local Governments for Sustainability carbon*n*® Climate Registry** | ICLEI Local Governments for Sustainability carbon*n*® Climate Registry (www.carbonn.org). (Data provided directly by ICLEI in June 2019). Individual targets and action plans for carbon*n* participants based on 2018 GPC Inventory responses. Although many ICLEI members report to CDP through the ICLEI-CDP unified reporting system, we supplemented our database with previously-collected data from the ICLEI carbon*n* Climate Registry if used in previous report editions. |
| **CDP Cities** | CDP. (2021). 2020 Full Cities Disclosure. Individual target and emissions data. Accessed February 2021 from: www.data.cdp.net. |
| **CDP 2020 Disclosure Survey** | CDP. (Provided directly from CDP in March 2021). *GHG emissions and action data for companies based on the 2020 responses*. |
| **Global Covenant of Mayors for Climate & Energy** | Global Covenant of Mayors for Climate & Energy. Membership data. Accessed February 2021 from www.globalcovenantofmayors.org. |
| **US Climate Alliance** | U.S. Climate Alliance. Accessed February 2021 from: https://www.usclimatealliance.org/state-climate-energy-policies.<br>Information from this source was supplemented through desk research of participants' climate action targets or plans. |
| **US Climate Mayors** | US Climate Mayors. Accessed February 2021 from: www.climatemayors.org and http://climatemayors.org/actions/climate-action-compendium/.<br>Information from this source was supplemented through desk research of participants' climate action targets or plans. |

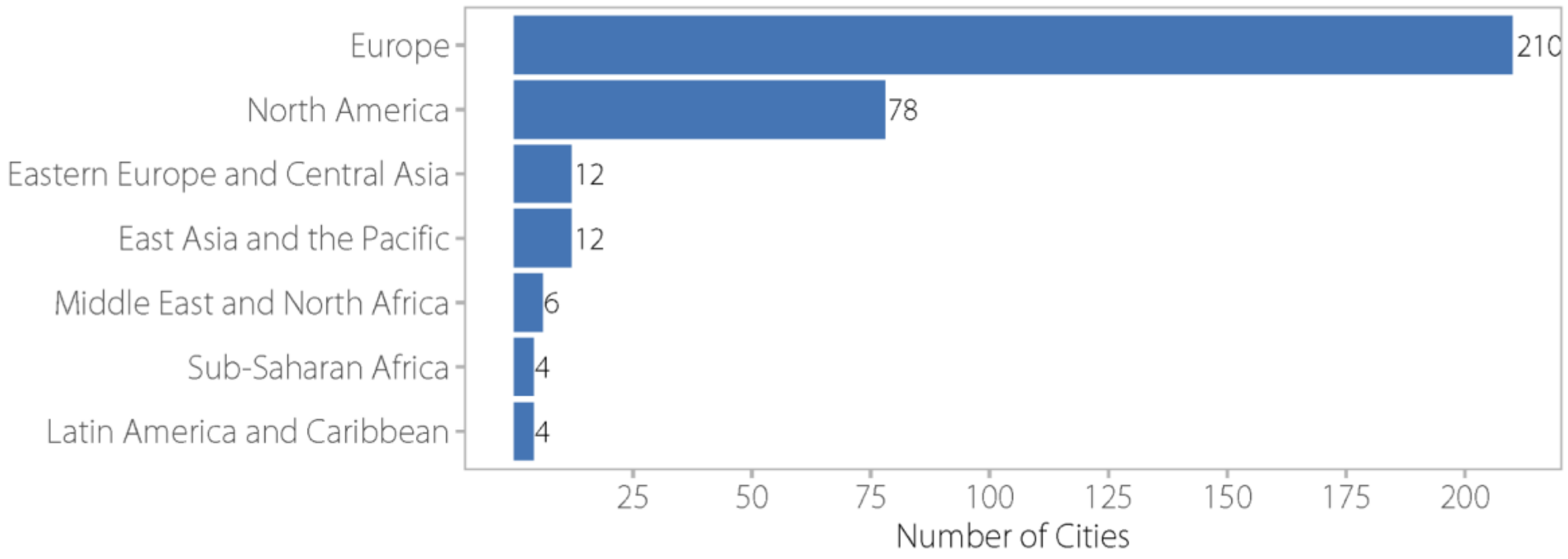

**Supplementary Figure 1**. Cities analyzed in the text document corpus, classified by World Bank Region.

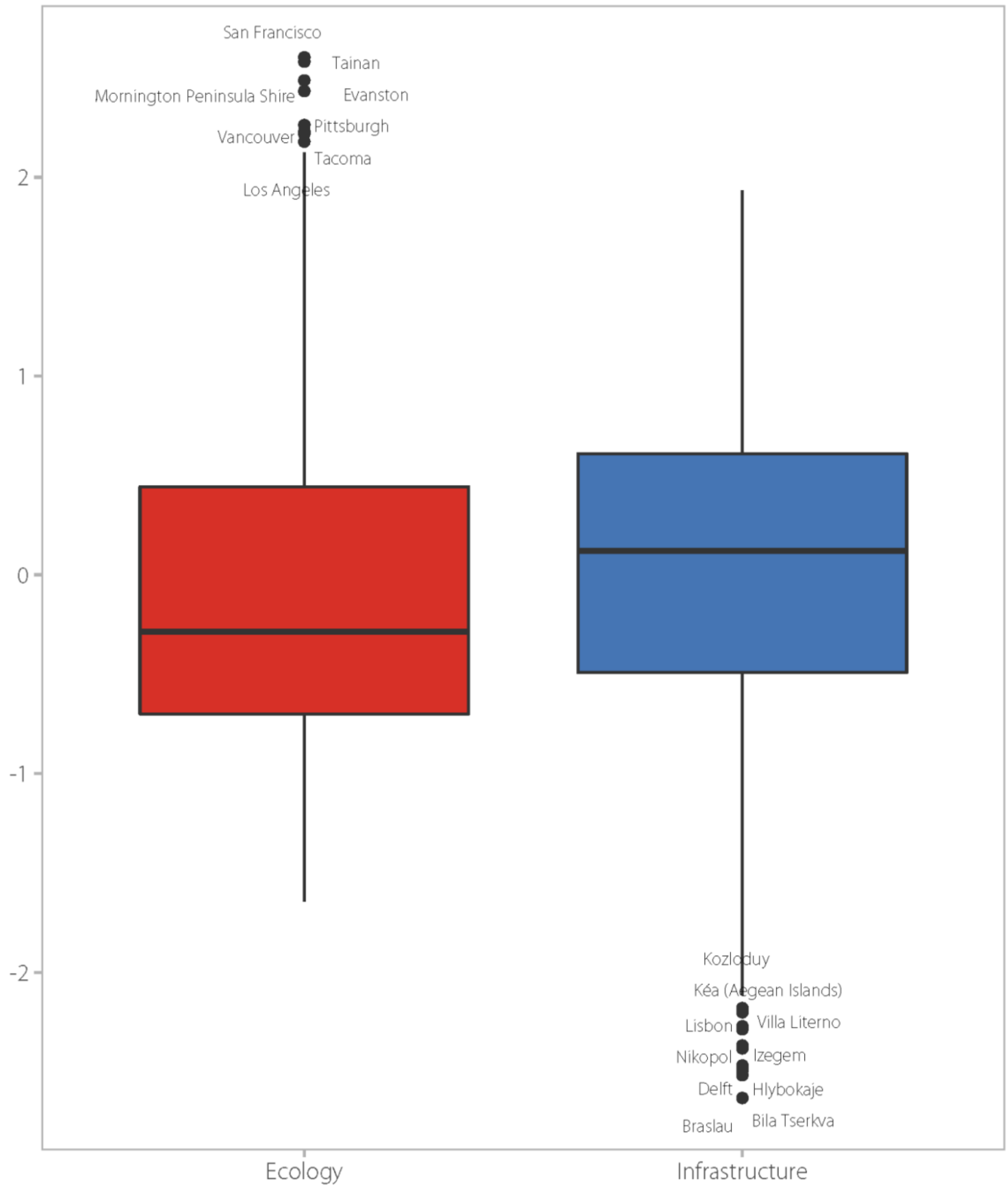

**Supplementary Figure 2**. Boxplot of factor scores on the two latent factors identified in cities' action plan that broadly emphasize Ecology and Infrastructure.

**Supplementary Table S2.** Performance metrics.

| | |
|---|---|
| f1-score | 0.812863 |
| precision | 0.840619 |
| recall | 0.811000 |

We report accuracy metrics ranging from 0 to 1, including binary f1-score, precision, and recall averaged over 50 out-of-sample test sets. We report out-of-sample accuracy on our logistic regression methodology of binary classification of cities based on net zero ambition. We evaluate by splitting the dataset into 50 train-test splits, training a model and evaluating on out of sample data using our regularization hyperparameter learned from cross-validation, and compute the average precision, recall, and f1 score. Because out of sample binary classification accuracy is the primary metric by which we evaluate logistic regression, these results indicate that the model we learn is robust to unseen data and therefore useful for interpretive analyses of the coefficients.

**Supplementary Table S3: Lexicon for key-term based topic analysis.**

| land use | offsets | transportation | heating | energy | pollution/ waste | building | climate impacts | industry |
|---|---|---|---|---|---|---|---|---|
| agriculture | offsets | transportation | btu | hydrogen | incinerator | interior | flooding | industrial |
| irrigation | offset | transit | heater | hydrocarbon | waste | build | inundation | industries |
| agricultural | offset purchase | transport | thermostat | megawatt | landfills | plumbing | levee | machinery |
| farming | emission offset | rail | therm | electrification | landfill | refurbishment | flood | industry |
| farmers | offset carbon | driverless | heaters | hydro | incineration | house | inundate | additive |
| livestock | carbon offset | highways | gallon | kwh | effluent | wall | downpour | chemical |
| farmland | negative emission | intercity | heating | electricity | sludge | tower | rainstorm | mining |
| farm | carbon credit | streetcar | boiler | kilowatt | disposal | floor | rainfall | mechanical |
| orchard | carbon capture | rideshare | cooling | vehicles | dispose | substation | storm | factory |
| permacul-ture | absorb carbon | ridesharing | heating | energy | recycling | homes | heat-wave | steel |
| rural | carbon sink | ridership | boiler | megawatt | composting | architec-tural | drought | metal |
| crop | remove carbon | carsharing | heat | electrifi-cation | garbage | architec-ture | katrina | materials |
| dairy | sequester | vanpool | tempera-ture | hydro | trash | facility | hurri-cane | incinera-tor |
| fisheries | carbon sequester | commuter | thermal | kwh | recycle | lighting | rain | freight |
| farmer | seques-tration | airport | btu | electric-ity | dumping | recon-struction | prepar-edness | tonnage |
| grower | carbon seques-tration | travel | heater | kilowatt | recyclable | residence | disaster | intermod-al |
| landowner | sequester carbon | vehicles | thermo-stat | cogene-ration | compost-able | work | seismic | manufac-ture |
| cattle | soil carbon | passenger | therm | biomass | recycled | home | tributary | operations |
| erosion | carbon credits | occupant | heaters | renew-able | biodegrad-able | dwelling | water-shed | ware-house |
| runoff | offset emission | commut-ing | ventila-tion | hydro-power | scrap | architect | water-course | goods |
| soil | | telecom-muting | snow-pack | renew-ables | pollution | architects | catch-ment | bulk |
| sediment | | tram | overheat | bioener-gy | polluted | retrofit | earth-quake | heavy |
| fertilizer | | bus | furnace | geother-mal | polluting | retrofits | devastat-ing | packaging |
| manure | | railway | hvac | biogas | smog | construc-tion | severe | manufac0t urer |
| aquaponic | | railroad | insula-tion | power | polluted | demoli-tion | tornado | facilities |
| land | | highway | insulate | photo-voltaic | pollutant | renova-tion | basin | inspector |
| commodity | | roadway | flow | solar | compost | building | impacts | inspection |
| drought | | taxis | steam | fuels | toxic | concrete | debris | manufac-turing |
| plants | | minibus | melt | fuel | methane | paving | drainage | supply chain |
| hydroponic | | taxi | circula-tion | electric | hazardous | infill | rainwa-ter | |
| harvesting | | roads | cool | wind | contami-nated | buildings | influx | |
| harvest | | parking | refriger-ation | turbine | contami-nant | structural | ground-water | |
| food | | train | stove | hydrocar-bon | contami-nate | design | hazard | |
| organics | | vehicle | dryer | coal | contamina-tion | subcon-tractor | wildfire | |
| organic | | routes | air | refinery | litter | contractor | burning | |
| pruning | | bicycle | electri-cal | radiation | dump | residen-tial | crisis | |
| planting | | commute | condi-tioning | biodiesel | sewage | multifam-ily | climatic | |
| landscapes | | carpool | condi-tioner | feedstock | sewer | nonresi-dential | survive | |
| landscape | | ferry | refriger-ator | ethanol | wastewater | mainte-nance | evacua-tion | |
| plant | | trucks | refriger-ant | biofuel | stormwater | structure | shelter | |
| vegetable | | driving | blower | fuel | plastic | construct | destruc-tive | |
| fertilizer | | cars | | kerosene | cleaning | builders | catas-trophic | |
| manure | | mileage | | propane | sulfur | built | potable | |
| acre | | mpg | | fuels | | facade | weather | |
| acreage | | engine | | nuclear | | exterior | forecast | |
| hectare | | exhaust | | combus-tion | | bridge | destruc-tion | |
| forestry | | tailpipe | | emission | | shingle | damage | |
| land | | motor | | biodiesel | | roofs | threat | |
| environ-mentally | | cars | | feedstock | | roofing | damag-ing | |
| greening | | metro | | ethanol | | roof | danger-ous | |
| livable | | bike | | biofuel | | housing | vulnera-ble | |
| liveable | | gasoline | | energy | | develop-ment | threaten | |
| biodiver-sity | | gas | | glazing | | durable | endanger | |
| steward-ship | | diesel | | combust | | builder | risk | |
| conservan-cy | | petrol | | hybrid | | installa-tion | harm | |
| reforesta-tion | | petroleum | | charging | | permit-ting | | |
| conserva-tion | | oil | | battery | | developm ents | | |
| deforesta-tion | | shuttle | | lpg | | | | |
| ozone | | flight | | fossil | | | | |
| biodiverse | | walking | | powered | | | | |
| estuary | | trip | | grid | | | | |
| woodland | | aviation | | efficien-cy | | | | |
| forest | | motorist | | utiliza-tion | | | | |
| riparian | | truck | | supply | | | | |
| prairie | | motorcycle | | generat-ing | | | | |
| floodplain | | station | | efficient | | | | |
| wetlands | | wheel | | generate | | | | |
| grassland | | subway | | generator | | | | |
| wetland | | bikeway | | genera-tion | | | | |
| ecosystem | | journeys | | pipeline | | | | |
| ecosystems | | driver | | sunlight | | | | |
| hydrologi-cal | | bicycling | | fluctua-tion | | | | |
| greening | | street | | surge | | | | |
| fauna | | destination | | | | | | |
| wildlife | | cyclist | | | | | | |
| vegetation | | car | | | | | | |
| habitat | | journey | | | | | | |
| pollinator | | vehicular | | | | | | |
| preserva-tion | | walk | | | | | | |
| ecological | | bikeshare | | | | | | |
| forestry | | automobile | | | | | | |
| woodland | | route | | | | | | |
| forest | | hauler | | | | | | |
| ecosystem | | scooter | | | | | | |
| ecosystems | | pedestrian | | | | | | |
| hydrologi-cal | | bicyclist | | | | | | |
| species | | cycling | | | | | | |
| specie | | biking | | | | | | |
| biodiverse | | downtown | | | | | | |
| reclamation | | auto | | | | | | |
| restoration | | gear | | | | | | |
| marine | | traffic | | | | | | |
| parks | | rooftop | | | | | | |
| greenspace | | drive | | | | | | |
| ozone | | carpooling | | | | | | |
| nature | | rider | | | | | | |
| animal | | road | | | | | | |
| extinction | | ride | | | | | | |
| botanical | | lane | | | | | | |
| parkland | | corridor | | | | | | |
| gardens | | walkway | | | | | | |
| bird | | sidewalk | | | | | | |
| desert | | ramp | | | | | | |
| trees | | tire | | | | | | |
| sea | | driveway | | | | | | |
| basin | | mile | | | | | | |
| waterway | | haul | | | | | | |
| river | | parking | | | | | | |
| ocean | | freight | | | | | | |
| marsh | | logistics | | | | | | |
| garden | | | | | | | | |
| gardening | | | | | | | | |
| ecologi-cally | | | | | | | | |
| grain | | | | | | | | |
| pesticide | | | | | | | | |
| waters | | | | | | | | |
| regenera-tive | | | | | | | | |
| creek | | | | | | | | |
| fruit | | | | | | | | |
| pest | | | | | | | | |
| grass | | | | | | | | |
| regenera-<br>tion | | | | | | | | |
| gardening | | | | | | | | |
| tree | | | | | | | | |
| lake | | | | | | | | |
| pond | | | | | | | | |
| aquatic | | | | | | | | |
| wild | | | | | | | | |
| fish | | | | | | | | |
| oceanic | | | | | | | | |
| insect | | | | | | | | |
| nutrient | | | | | | | | |
| fishing | | | | | | | | |
| hillside | | | | | | | | |
| shore | | | | | | | | |
| shoreline | | | | | | | | |
| gravel | | | | | | | | |
| seed | | | | | | | | |
| weed | | | | | | | | |
| habitat | | | | | | | | |
| parks | | | | | | | | |
| parks | | | | | | | | |
| parcel | | | | | | | | |